\input harvmac.tex
\hfuzz 15pt
%\draftmode
\input amssym.def
\input amssym.tex
\input epsf\def\tfig#1{{
\xdef#1{Fig.\thinspace\the\figno}}Fig.\thinspace\the\figno
\global\advance\figno by1}

%%%%%%%%%%%%%%%%%%%DEFINITIONS%%%%%

\input epsf

%

% font definitions

%%%%%%%%%%%%%%%%%%%%%%%%%%%%

\def\bp{\bar\partial}

%%%%%%%%%GREEK LETTERS%%%%%%%%%%%%%%%%
\def\a{\alpha}
\def\b{\beta}
\def\g{\gamma}
\def\d{\delta}

\def\vp{\varphi}
\def\G{\Gamma}

\def\ov{\over}
\def\o{\omega }
\def\bps{\bar\psi}

\def\no{\noindent}
%%%%%%%%%%%%%%%%%%%ELLIPTIC%%%%%%%%%%%%%

%%%%%%%%%%%%%%%Im%%%%Re%%%%%%%%%%%%%%

%%%%%%%%%%%%%%%%%%%%% Calligraphic letters %%%%%
 %

\def\[{\left[}
\def\]{\right]}
\def\({\left(}
\def\){\right)}
\def\<{\left\langle\,}
\def\>{\,\right\rangle}
\def\half{ {1\over 2} }
%%%%%%%%%%%%%%%%%%%%%%%%
\def\d{\partial}

\def\inv{^{-1}}

 \def\frac#1#2{ {{\textstyle{#1\over#2}}}}
\def\inv{^{\raise.15ex\hbox{${\scriptscriptstyle -}$}\kern-.05em 1}}

 \def\IP{\relax{\rm I\kern-.18em P}}

%% G changed to C
% \def\GL{ C^{\rm Liou}}
%  \def\GM{ C^{\rm Matt }}

%

%\def\za{\alpha}

% \def\zg{\gamma} \def\zd{\delta}
%\def\ze{\varepsilon} \def\zf{\phi} \def\zv{\varphi} \def\zh{\chi}
%\def\zk{\kappa}
%%\def\zl{\lambda}
%\def\zm{\mu} \def\zn{\nu}
%\def\zo{\omega} \def\zp{\pi} \def\zr{\rho}

%\def\zt{\tau} \def\zz{\dzeta}\def\zi{\eta}

\def\la{\langle} \def\ra{\rangle}

\def\dC{C\kern-6.5pt I}

              \def\CL{{\cal L}}

%

%%%%%%%%%%%%%%%%%%%

%%%%%%%Russian fonts%%%%%%%5

\chardef\tempcat=\the\catcode`\@ \catcode`\@=11
\def\cyracc{\def\u##1{\if \i##1\accent"24 i%
    \else \accent"24 ##1\fi }}
\newfam\cyrfam

%%%%% REFS %%%%%%%%%%

%%%%%%

\def\np#1#2#3{{Nucl. Phys.} {\bf B#1} (#2) #3}
\def\pl#1#2#3{{Phys. Lett. }{\bf B#1} (#2) #3}

\def\physrev#1#2#3{{Phys. Rev.} {\bf D#1} (#2) #3}

\def\ijmp#1#2#3{{Int. J. Mod. Phys.} {\bf #1} (#2) #3}

\def\tmatp#1#2#3{{Theor. Math. Phys.} {\bf #1} (#2) #3}
\def\hep#1{{hep-th/}#1}

\def\encadremath#1{\vbox{\hrule\hbox{\vrule\kern8pt\vbox{\kern8pt
 \hbox{$\displaystyle #1$}\kern8pt}
 \kern8pt\vrule}\hrule}}

\def\gpl{G^+}
\def\gmi{G^-}
\def\gph{G^+_{-\half}}
\def\gmh{G^-_{-\half}}

%%%%%%%%%%%%%%%%%%%%%%%%%%%%%%%%%%%%%%%%%%%%%%%%%%%%%%%%%%%%%%%

\def\hepth#1{{arXiv:hep-th/}#1}

\def\np#1#2#3{{Nucl. Phys.} {\bf B#1} (#2) #3}
\def\pl#1#2#3{{Phys. Lett. }{\bf B#1} (#2) #3}

\def\physrev#1#2#3{{Phys. Rev.} {\bf D#1} (#2) #3}

\def\ijmp#1#2#3{{Int. J. Mod. Phys.} {\bf #1} (#2) #3}

\def\tmatp#1#2#3{{Theor. Math. Phys.} {\bf #1} (#2) #3}
%\def\hepth#1{{e-Print Archive:  hep-th/}#1}
%\def\hepth#1{{hep-th/}#1}

%%%%%%%%%%%%%%%%%%%%%%%%%%%%%%%%%%%%%%%%%%%%%%%%%%%%%%%

\lref\al{Al. Zamolodchikov, \ijmp{A19[S2]}{2004}510.;
\hepth{0312279}.}

\lref\grav{Al. Zamolodchikov, \tmatp{142}{2005}183;
\hepth{0505063}.}

\lref\gravo{A. Belavin and Al. Zamolodchikov, \tmatp{147}{2006}339;
           \hepth{0510214}.}

\lref\gravtw{A. Belavin and V. Belavin, J. Phys.{\bf A42}: 304003 (2009);
\hepth{08101023}.}

\lref\gravth{V. Belavin, \tmatp{161}{2009}1361; \hepth{09024407}.}

\lref\slft{A. Belavin and Al. Zamolodchikov, JETP Lett. {\bf 84} (2006)
418; \hepth{0610316}.}

\lref\wzw{G. Bertoldi and G. Giribet, \np{701}{2004}481;
\hep{0405094}.}

\lref\wzwo{G. Bertoldi, S. Bolognesi, G. Giribet, M. Matone, and Yu.
Nakayama, \np{709}{2005}522; \hep{0409227}.}

\lref\heb{A. Belavin and V. Belavin, JHEP {\bf 1002} (2010) 010;
\hepth{09114597}.}

\lref\ntl{A. Giveon and D. Kutasov, JHEP {\bf 9910} (1999) 034;
\hep{9909110}.}

\lref\ntlo{A. Giveon, D. Kutasov, and A. Schwimmer, \np{615}{2001}133;
\hep{0106005}.}

\lref\ntltw{K. Hori and A. Kapustin, JHEP {\bf 0108} (2001) 045;
\hep{0104202}.}

\lref\dual{C. Ahn, C. Kim, C. Rim, and M. Stanishkov,
\physrev{69}{2004}106011; \hep{0210208}.}

\lref\dualo{Y. Nakayama, Int. J. Mod. Phys. {\bf A19} (2004) 2771; \hep{0402009}.}

\lref\ntun{W. Boucher, D. Friedan, and A. Kent, \pl{172}{1986}316.}

\lref\ntuno{S. Nam, \pl{172}{1986}323.}

\lref\ntuntw{G. Mussardo, G. Sotkov, and M. Stanishkov,
\ijmp{A4}{1989}1135.}

\lref\reps{C. Ahn, M. Stanishkov, and M. Yamamoto, JHEP {\bf 0407}
(2004) 057; \hepth{0405274}.}

\lref\repso{K. Hosomichi, JHEP {\bf 0612} (2006) 061;
\hepth{0408172}.}

\lref\zzam{A. Zamolodchikov and Al. Zamolodchikov, {\it Liouville field
            theory on a pseudosphere}; \hep{0101182}.}

\lref\aybc{C. Ahn and M. Yamamoto, \physrev{69}{2004}026007;
\hepth{0310046}.}

%\vskip 1.5cm
%\noindent

%${\bf N=0}$ case

%$D_{mn}$ = singular vectors at level mn

%Definition of $B_{mn}$:

%$$
%D_{mn}\bar D_{mn}V'_{mn}=B_{mn}V_{m,-n}
%$$

%(remember $\Delta_{mn}+mn=\Delta_{m,-n}$)

%where the "logarithmic field" is

%$$
%V'_\alpha={1\over 2}\partial_\alpha V_\alpha
%$$

%For degenerate fields:

%$$
%V'_{mn}=V'_\alpha|_{\alpha=\alpha_{mn}}
%$$

%Definition of $r_{mn}$:

%$$
%<\alpha|D_{mn}^\dagger
%D_{mn}|\alpha>=(\alpha-\alpha_{mn})r_{mn}+o((\alpha-\alpha_{mn})^2)
%$$

%Result:

%$$
%r_{mn}=2\prod_{k=1-n}^n\prod_{l=1-m}^m (lb^{-1}+kb)
%$$

%($(k,l)\nonequal (0,0)$ )

%Then:

%$$
%r_{mn}=2 {U_{p,q}(\tilde\alpha_{m,n})\over U_{p,q}(\alpha_{mn})}
%B_{m,n}
%$$
%\eject

\rightline{In memory of Alyosha Zamolodchikov}
\vskip -1cm
\overfullrule=0pt
 %%%
\Title{\vbox{\baselineskip12pt\hbox {}\hbox{}}} {\vbox{\centerline
 {Higher Equations of Motion in $N=2$ Superconformal }
%  {3-point open non-critical string correlators of ZZ type}
 \bigskip
 \centerline{Liouville Field Theory  } % ${}^*$}
  \vskip2pt
}} \centerline{Changrim Ahn$^{a,b,}$\foot{ahn@ewha.ac.kr}, Marian
Stanishkov$^{b,}$\foot{marian@inrne.bas.bg; {\it On leave of absence
from INRNE, BAS, Bulgaria}} and Mihail
Stoilov$^{c,}$\foot{mstoilov@inrne.bas.bg} }

 \vskip 1cm

 \centerline{ \vbox{\baselineskip12pt\hbox
{\it $^{a)}$Department of Physics, Ewha Womans University, Seoul
120-750, S. Korea}
 }}
\centerline{ \vbox{\baselineskip12pt\hbox {\it $^{b)}$Institute for
the Early Universe, Ewha Womans University, Seoul 120-750, S. Korea}
 }}
\centerline{  {\it  $^{c)}$Institute for Nuclear Research and
Nuclear Energy (INRNE), } }
 \centerline{ \vbox{\baselineskip12pt\hbox
 {\it Bulgarian Academy of Sciences (BAS), 1784 Sofia, Bulgaria}
 }}

 %%%%%%%%%%%%%%%%%%%%%%%%%%%%%%%%%

\vskip 1.5cm

\centerline{ Abstract} \vskip.5cm \noindent \vbox{\baselineskip=15pt
We present an infinite set of higher equations of motion in $N=2$
supersymmetric Liouville field theory. They are in one to one
correspondence with the degenerate representations and are
enumerated in addition to the $U(1)$ charge $\o$ by the positive
integers $m$ or $(m,n)$ respectively. We check that in the classical
limit these equations hold as relations among  the classical
fields.}

\Date{}
\vfill \eject

In ref.\al\ it has been shown that in the Liouville field theory
(LFT) an infinite set of relations holds for quantum operators.
These equations relate different basic Liouville primary fields
$V_\a(z)$ ($V_\a$ can be thought of as normal ordered exponential
field ${\rm exp}(\a\phi)$ of the basic Liouville field $\phi$). They
are parameterized by a pair of positive integers $(m,n)$ and are
called conventionally ``higher equations of motion'' (HEM), because
the first one $(1,1)$ coincides with the usual Liouville equation of
motion. The equations are derived on the basis of a conjecture of
the vanishing of all singular vectors, imposed by the requirement of
irreducibility of the corresponding representation. They are easily
verified in the classical LFT. Higher equations turn out to be
useful in practical calculations. In particular, in
\refs{\grav\gravo\gravtw-\gravth}, they were used to derive general
four-point correlation function in the minimal Liouville gravity.

Similar operator valued relations have been found also for $N=1$
supersymmetric Liouville field theory (SLFT) \slft\ and for $SL(2,R)$
Wess-Zumino-Novikov-Witten
model \refs{\wzw,\wzwo} . Recently it was shown in \heb\ that such
relations hold for the boundary operators in the LFT with conformal
boundary.

It is the purpose of this note to reveal a similar
set of higher equations of motion in $N=2$ SLFT.
The $N=2$ SLFT has a wide variety of
applications in string theory \refs{\ntl\ntlo-\ntltw} . This theory
is quite interesting because of the fact that it has actually few
properties in common with the $N=0,1$ SLFTs. For example, unlike the
Liouville theories with less supersymmetry, the $N=2$ SLFT does not have
a simple strong-weak coupling duality. In fact, under the change
$b\to 1/b$ of the coupling constant, the $N=2$ SLFT flows to another
$N=2$ supersymmetric theory as proposed in \refs{\dual,\dualo}.
Another important difference between the $N=2$ SLFT and the $N=0,1$
SLFTs is the spectrum of the degenerate representations
\refs{\ntun\ntuno-\ntuntw} (see also \refs{\reps,\repso} ). We will
show below that the $N=2$ SLFT still possesses
higher equations of motion despite these differences.

\vskip.5cm
\noindent {\bf $N=2$ SLFT} 

The $N=2$ SLFT is based on the
Lagrangian:
\eqn\act{ \eqalign{ \CL &= {1\ov
{2\pi}}\(\d\phi^-\bp\phi^+ +\d\phi^+\bp\phi^- +\psi^-\bp\psi^+
+\psi^+\bp\psi^- +\bps^-\d\bps^+ +\bps^+\d\bps^- \)+\cr
 &+i\mu b^2\psi^-\bps^-e^{b\phi^+}
+i\mu b^2\psi^+\bps^+e^{b\phi^-}+\pi\mu^2 b^2 e^{b\phi^+ +b\phi^-}\
} } where $(\phi^\pm, \psi^\mp)$ are the components of a chiral
$N=2$ supermultiplet, $b$ is the coupling constant and $\mu$ is the
cosmological constant. It is invariant under the $N=2$
superconformal algebra: \eqn\ntwo{ \eqalign{ [L_m,L_n]
&=(m-n)L_{m+n}+{c\ov 12}(m^3-m)\delta_{m+n},\cr [L_m,G^{\pm}_r]
&=\left({m\ov 2}-r\right)G^{\pm}_{m+r},\hskip1cm [J_n,G^{\pm}_r]=\pm
G^{\pm}_{n+r},\cr \{G^+_r,G^-_s\} &=2L_{r+s}+(r-s)J_{r+s}+{c\ov
3}\left(r^2-{1\ov 4}\right)\delta_{r+s},\cr [L_m,J_n] &=-nJ_{m+n},\hskip1cm
[J_m,J_n]={c\ov 3}\delta_{m+n},\ } } where $L_m, G^{\pm}_r$ and $J_n$
are the modes of the corresponding conserved currents, the
stress-energy tensor $T(z)$, the super-current $G(z)$ and the $U(1)$
current $J(z)$, and the central charge is:
$$
c=3+{6\ov b^2}.
$$
These are the left handed generators, there are in addition the
right handed ones $\bar L_n$, $\bar J_n$, $\bar G^{\pm}_r$ closing
the same algebra. 

The basic objects are the primary fields (normal
ordered exponents):
$$
N_{\a,\bar\a}=e^{\a\phi^++\bar\a\phi^-},
$$
the corresponding states being annihilated by the positive modes.
These are the primary fields in the Neveu-Schwartz (NS) sector with
$r$, $s$ in \ntwo\ half-integer. There are in addition also Ramond
($r$, $s$ - integer) primary fields $R_{\a,\bar\a}$ but we will not
be concerned with them in this paper. The conformal dimension and
the $U(1)$ charge of the primary fields are: \eqn\delta{
\Delta_{\a,\bar\a}=-\a\bar\a+{1\ov {2b}}(\a+\bar\a),\hskip1cm
\o={1\ov b}(\a-\bar\a) . }

Among the primary fields there is a series of degenerate
fields of the $N=2$ SLFT. They are characterized by the fact that at
certain level of the corresponding conformal family a new primary
field (i.e. annihilated by all positive modes) appears. Such fields
can be divided  in three classes (see e.g. \reps\ ). 

Class I degenerate fields are given by \eqn\clone{ \eqalign{ \ &\hskip2cm
N_{m,n}^\o=N_{\a_{m,n}^\o,\bar\a_{m,n}^\o},\cr \a_{m,n}^\o &={1-m\ov
{2b}}+(\o-n){b\ov 2},\hskip1cm
\bar\a_{m,n}^\o={1-m\ov{2b}}-(\o+n){b\ov 2}\ } } $m,n$ are positive
integers. $N_{m,n}^\o$ is degenerate at level $mn$ and relative
$U(1)$ charge zero. The irreducibility of the corresponding
representations is assured by imposing the null-vector condition
$D_{m,n}^\o N_{m,n}^\o=0$, $\bar D_{m,n}^\o N_{m,n}^\o=0$, where
$D_{m,n}^\o$ is a polynomial of the generators in \ntwo\ of degree
$mn$ and has $U(1)$ charge zero. It is normalized by choosing the
coefficient in front of $(L_{-1})^{mn}$ to be $1$. 
Let us give some
examples of the corresponding null-operators: \eqn\nullmn{ \eqalign{
D^\o_{1,1} &=L_{-1}-\half b^2(1+\o)J_{-1}+{1\ov \o-1}\gph\gmh, \cr
D_{1,2}^\o &=L^2_{-1}+b^2L_{-2}-b^2(1+\o)L_{-1}J_{-1}+
{b^2\ov{2}}\left(1+\o-b^2(2+\o)\right)J_{-2}+\cr &+{b^4\ov 4}\o(\o+2)J^2_{-1}+{2\ov
{\o-2}}L_{-1}\gph\gmh-{b^2\o\ov {\o-2}}J_{-1}\gph\gmh-\cr &-{b^2\ov{2}}
\gph\gmi_{-{3\ov 2}}+{b^2\ov{2}}{{\o+2}\ov {\o-2}}\gpl_{-{3\ov
2}}\gmh, \cr D_{2,1}^\o &=L^2_{-1}+{1\ov
b^2}L_{-2}-b^2(1+\o)L_{-1}J_{-1}+\half \left(b^2(1+\o)-\o-2\right)J_{-2} +\cr
&+{1\ov 4}\left(b^4(\o+1)^2-1\right)J^2_{-1}+{2b^4\o\ov
b^4(\o-1)^2-1}L_{-1}\gph\gmh-\cr &-{{b^2+b^6(\o^2-1)}\ov
b^4(\o-1)^2-1}J_{-1}\gph\gmh -{{b^4(\o+1)+b^2-2}\ov
2+2b^2(\o-1)}\gph\gmi_{-{3\ov 2}}+\cr
&+{{2-b^2+b^4(\o-1)\left(1+b^2(\o+1)\right)}\ov 2(b^4(\o-1)^2-1)}\gpl_{-{3\ov
2}}\gmh .\ }} 

The second class of degenerate fields is
denoted by $N_m^\o$ and comes in two subclasses IIA and IIB:
\eqn\cltwo{ \eqalign{ \rm{class} &\ {\rm IIA}:\hskip1cm
N_m^\o=N_{\a_m^\o,\bar\a_m^0}\hskip1cm \o>0\cr \rm{class} &\
{\rm IIB}:\hskip1cm N_m^\o=N_{\a_m^0,\bar\a_m^\o}\hskip1cm \o<0\ . }}
where \eqn\atwo{ \a_m^\o={1-m\ov {2b}}+\o b,\hskip1cm
\bar\a_m^\o={1-m\ov {2b}}-\o b. } Here $m$ is an odd positive
integer number and the level of degeneracy of $N_m^\o$ is ${m\ov
2}$, relative charge $\pm 1$. In this case the operator $D_m^\o$ is
a polynomial of ``degree'' $m/2$, the coefficient in front of
$L_{-1}^{{{m-1}\ov 2}}G^\pm_{-\half}$ is chosen to be $1$. Analogously
to the class I we have to impose $D_m^\o N_m^\o=\bar D_m^\o
N_m^\o=0$. Here are the first examples for class IIA fields:
\eqn\nullm{ \eqalign{ D_1^\o &=G^+_{-\half},\cr D_3^\o
&=L_{-1}G^+_{-\half}-J_{-1}G^+_{-\half}+\left({2\ov {b^2}}-\o\right)G^+_{-{3\ov
2}},\cr D_5^\o &=L_{-1}^2G^+_{-\half}+ \left({4\ov
{b^2}}-\o-1\right)L_{-2}G^+_{-\half}-3L_{-1}J_{-1}\gpl_{-\half}+2J_{-1}^2\gph+\cr
&+\left({5\ov 2}-{6\ov {b^2}}+{3\ov 2}\o\right)J_{-2}\gpl_{-\half}+\left(1+{6\ov
{b^2}}-2\o\right)L_{-1}\gpl_{-{3\ov 2}}+4\left(\o-{3\ov
{b^2}}\right)J_{-1}\gpl_{-{3\ov 2}}-\cr &-{\half}\gpl_{-{3\ov
2}}\gph\gmi_{-\half} + \left({24\ov {b^4}}-{14\o\ov
{b^2}}+2\o^2-1\right)\gpl_{-{5\ov2}}.\ }} The null-operators for class IIB
fields are obtained from \nullm\ by changing $G^\pm\to G^\mp$ and
$\o\to-\o$.

A special case of Class IIA (B) fields are the chiral
(antichiral) fields with $m=1$. The Class II fields having $U(1)$
charge zero are classified in a separate Class III fields. The simplest
$m=1$ field here represents the identity operator.

\vskip.5cm \noindent {\bf Norms of the null-states} 

Let us now
consider, for a further use, the norms of the states created by
applying the null-operators on primary states $|\a\rangle$. As explained
above, such sates should vanish at $\a=\a_M^\o$. Taking the first
terms in the corresponding Taylor expansion, we define: \eqn\defr{
\eqalign{
 r_M^\o &=\partial_\alpha\langle\a,\bar\a|D^{\o\dagger}_M
 D_M^\o|\a,\bar\a\rangle|_{\a=\a_M^\o,\bar\a=\bar\a_M^\o},\cr
\bar r_M^\o &=\partial_{\bar\alpha}\langle\a,\bar\a|D^{\o\dagger}_M
D_M^\o|\a,\bar\a\rangle|_{\a=\a_M^\o,\bar\a=\bar\a_M^\o}\ }} for both
classes of representations, $M=m$ or ($m,n$), where $D_M^\o$ is the
corresponding null-operator and $D^{\o\dagger}_M$ is defined as
usual through $L^\dagger_n=L_{-n}$, $J^\dagger_n=J_{-n}$,
$(G^{\pm}_r)^\dagger=G^\mp_{-r}$. 

One can compute ``by hand'' the
first few $r$'s. With the use of the explicit form of the null-operators
\nullmn\ we find for the class I fields:
$$
\eqalign{ r_{1,1}^\o &={1\over b}{(1+b^2)(1+\o)\over (-1+\o)},\cr
r_{1,2}^\o &={-2\over b}{(1-b^2)(1+b^2)(1+2b^2)(2+\o)\over
(-2+\o)},\cr r_{1,3}^\o &={12\over
b}{(1-2b^2)(1-b^2)(1+b^2)(1+2b^2)(1+3b^2)(3+\o)\over (-3+\o)},\cr
r_{2,1}^\o &={2\over
b^5}{(1-b^2)(1+b^2)(2+b^2)(-1+b^2+b^2\o)(1+b^2+b^2\o)\over
(-1-b^2+b^2\o)(1-b^2+b^2\o)},\cr r_{3,1}^\o &={12\over
b^9}{(2-b^2)(1-b^2)(1+b^2)(2+b^2)(3+b^2)(1+\o)(-2+b^2+b^2\o)(2+b^2+b^2\o)\over
(-1+\o)(-2-b^2+b^2\o)(2-b^2+b^2\o)}\ }
$$
and $\bar r_{m,n}^\o=r_{m,n}^\o$  for all the examples above. Based
on these expression we propose for the general form of $r_{m,n}^\o$:
\eqn\rmno{ r_{m,n}^\o=\bar
r_{m,n}^\o=\prod_{l=1-m}^m\prod_{k=1-n}^n\left({l\over b}+kb\right)
\prod_{l=1-m,\ {\rm mod}\ 2}^{m-1}\left({{l-(n+\o)b^2}\over
l+(n-\o)b^2}\right). } 
Similarly, from \nullm\
we have for the class IIA:
$$
{\eqalign{ \bar r_1^\omega &=2\left({1\over b}-\omega b\right),\cr \bar r_3^\o
&={2\over b^5}(2-b^2\o)(3-b^2\o)(2-b^2-b^2\o),\cr \bar r_5^\o
&={8\over
b^9}(3-b^2\o)(4-b^2\o)(5-b^2\o)(3-b^2-b^2\o)(4-b^2-b^2\o),\cr \bar
r_7^\o &={72\over
b^{13}}(4-b^2\o)(5-b^2\o)(6-b^2\o)(7-b^2\o)(4-b^2-b^2\o)(5-b^2-b^2\o)(6-b^2-b^2\o),\cr
r_m^\o &=0,\quad m=1,3,5,7.\ }}
$$

These expressions can be fitted in a general form of $r_m^\o$ and
$\bar r_m^\o$:
\eqn\rmo{ \eqalign{
r_m^\o &= 0,\cr
 \bar r_m^\o &=2\G^2\left({{m+1}\over 2}\right) b^{1-m}
\prod_{l={{m+1}\over 2}}^{m}\left({l\over b}-b\o\right) \prod_{l={{m+1}\over
2}}^{m-1} \left({l\over b}-b(\o+1)\right).\ } } For the class IIB fields one
obtains $\bar r^\o_m=0$ and $r_m^\o$ is as $\bar r_m^\o$ in \rmo\
with the change $\o\to -\o$.

\vskip.5cm \noindent {\bf Logarithmic fields and HEM} 

Let us now
introduce the so called logarithmic fields. They are defined as:
$$
N'_{\a,\bar\a}=\d_\a N_{\a,\bar\a},\hskip1cm \bar
N'_{\a,\bar\a}=\d_{\bar\a}N_{\a,\bar\a}.
$$
One can introduce also the logarithmic primary fields corresponding
to degenerate fields by:
\eqn\log{
{N'}_M^{\o}
=N'_{\a,\bar\a} |_{\a=\a_M^\o,\bar\a=\bar\a_M^\o},\hskip1cm \bar {N'}_M^{\o}
=\bar N'_{\a,\bar\a}|_{\a=\a_M^\o,\bar\a=\bar\a_M^\o} }
where $M$
is ($m,n$) for class I and $M$ is $m$ for class II fields
respectively. The basic statement about the fields \log\ is that
\eqn\logpr{ \tilde N_M^\o=\bar D_M^\o D_M^\o {N'}^{\o}_M ,\hskip.5cm
\tilde{\bar N}_M^\o=\bar
 D_M^\o D_M^\o \bar {N'}^{\o}_M }
with $D_M^\o$, $\bar D_M^\o$ as in \nullmn , \nullm\ are again
primary. The proof of this statement goes along the same lines as
for $N=0,1$ SLFT (\al ,\slft) so we will not repeat it here.

Comparing the dimension and $U(1)$ charge for class I
fields: $\tilde\Delta_{m,n}= \Delta_{m,n}+mn,\ \tilde\o=\o$ we
conclude that the fields \logpr\ are proportional to $N_{m,-n}^\o$.
Thus, we arrive at the higher equations of motion (HEM) for the
class I fields:
\eqn\hemmn{ \bar D_{m,n}^\o D_{m,n}^\o
{N'}^{\o}_{m,n}=B_{m,n}^\o N_{m,-n}^\o,\qquad \bar D_{m,n}^\o D_{m,n}^\o
\bar {N'}^{\o}_{m,n}=\bar B_{m,n}^\o N_{m,-n}^\o .\ }

For class IIA (B) the dimension of the resulting primaries
in \logpr\ is $\tilde\Delta_m^\o=\Delta_m^\o+{m\ov 2}$, the $U(1)$
charges are $\tilde\o=\o+1$ ($\tilde\o=\o-1$) respectively, and the
HEMs in this case are:
\eqn\hemm{ \bar D_{m}^\o D_{m}^\o
{N'}^{\o}_{m} =B_{m}^\o N_{m}^{\o\pm 1},\qquad \bar D_{m}^\o D_{m}^\o
\bar {N'}^{\o}_{m} =\bar B_{m}^\o N_{m}^{\o\pm 1}.\ }
Computation of
$B_{m,n}^\o$ ($\bar B_{m,n}^\o$) and $B_m^\o$ ($\bar B_m^\o$) is the
final goal of this note. HEMs \hemmn\ and \hemm\ are to be
understood in an operator sense, i.e. they should hold for any
correlation function. Here we will insert them into the simplest
one-point function on the so called Poincar${\rm\acute{e}}$ disk
\zzam . In this case we have:
$$
\la B_1|\bar D_M^\o D_M^\o N^{'\o}_M\ra =\la B_1|\tilde
N_M^\o\ra,\qquad \la B_1|\bar D_M^\o D_M^\o \bar N^{'\o}_M\ra =\la
B_1|\tilde{\bar N}_M^\o\ra .
$$
The boundary state $\la B_1|$ corresponds to the identity boundary
conditions on the Poincar${\rm\acute{e}}$ disc. It enjoys $N=2$
superconformal invariance:
$$
\la B_1|\bar G^{\pm}_r =-i\la B_1|G^\mp_{-r}= -i\la
B_1|(G^{\pm}_{r})^\dagger,\quad \la B_1|\bar L_n,=\la
B_1|(L_{n})^\dagger,\quad
\la B_1|\bar J_n =\la
B_1|(J_{n})^\dagger.
$$
(so called A-type boundary conditions, see e.g. \aybc). 

With the definition of $r$'s in
\defr\ the HEMs \hemmn\ and \hemm\ take the form: \eqn\rbumn{
\eqalign{ r_{m,n}^\o U_1(m,n;\o) &=B_{m,n}^\o U_1(m,-n;\o),\cr \bar
r_{m,n}^\o U_1(m,n;\o) &=\bar B_{m,n}^\o U_1(m,-n;\o)\ }} for class
I, and \eqn\rbum{ \eqalign{ r_m^\o U_1(m,\o) &=iB_m^\o U_1(m,\o\pm
1),\cr \bar r_m^\o U_1(m,\o) &=i\bar B_m^\o U_1(m,\o\pm 1)\ }} for
class II. Here $U_1$ is the one-point function for ``identity
boundary conditions'' of the corresponding field. In \rbum\ the factor $i$'s
appear because the class II null-operators are fermionic, and $+$
($-$) refers to class IIA (IIB).

The one-point function on the Poincar${\rm\acute{e}}$ disk
for identity boundary conditions in $N=2$ SLFT was obtained in
\reps\ and has a general form:
$$
U_1(\a,\bar\a)= \G (b^{-2})(\pi\mu)^{-{1\ov b}(\a+\bar\a)}
{\G(1-\a b)\G(1-\bar\a b)\over\G(-{{\a+\bar\a}\over b}+{1\over
b^2})\G(2-b(\a+\bar\a))}.
$$
With the specific values \clone\ the ratio of one-point functions of
class I fields then is:
$$\eqalign{
{U_1(m,n;\o)\over U_1(m,-n;\o)}= &(\pi\mu)^{2n}{\g(1+m-nb^2)\over
\prod_{k=-n}^{n-1}({m\over b^2}+k) \prod_{l=-m}^{m}(l+nb^2)}
{\g({{1-m}\over 2}+(n-\o){b^2\over 2})\over \g({{1-m}\over
2}-(n+\o){b^2\over 2})}\times\cr &\times
\prod_{l=1-m,\ {\rm mod}\ 2}^{m-1}\left({{l+(n-\o)b^2}\over
l-(n+\o)b^2}\right)\ }
$$
and for the HEM coefficient we obtain: \eqn\bmno{ \eqalign{
B_{m,n}^\o &=\bar B_{m,n}^\o=r_{m,n}^\o {U_1(m,n;\o)\over U_1(m,-n;\o)}\cr
&=(\pi\mu)^{2n}b^{1+2n-2m}\g(m-nb^2) { \g({{1-m}\over
2}+(n-\o){b^2\over 2})\over \g({{1-m}\over 2}-(n+\o){b^2\over 2})}
\prod_{l=1-m}^{m-1}\prod_{k=1-n}^{n-1}\left({l\over b}+k b\right), }}
where we impose that $(k,l)= (0,0)$ is excluded in the product.

Analogously for class IIA fields:
$$
{U_1(m,\o)\over U_1(m,\o+1)}=\pi\mu b\ { \prod_{l={{m+1}\over
2}}^{m-1}({l\over b}-b(\o+1))\over \prod_{l={{m+1}\over
2}}^m({l\over b}-b\o)}
$$
and
\eqn\bmo{ \eqalign{ B_m^\o &=0,\cr \bar B_m^\o &=-i\bar
r_m^\o{U_1(m,\o)\over U_1(m,\o+1)}=-2\pi i\mu
b^{2-m}\G^2\left({{m+1}\over 2}\right)\prod_{l={{m+1}\over 2}}^{m-1}\left({l\over
b}-b(\o+1)\right)^2.\
 }}
For class IIB $B$ and $\bar B$ are exchanged and $\o$ is replaced by
$-\o$. Equalities \bmno\ and \bmo\ are the main results of this
paper.

\vskip.5cm \noindent {\bf Classical limit} 

In the classical limit
$b\to 0$: $b\phi\to\vp$, $\b\psi\to\psi$,
 $\pi\mu b^2\to M$ the Lagrangian $\CL\to {1\ov {2\pi
b^2}}\CL$. The corresponding equations of motion are given by
\eqn\em{
\eqalign{ \bp\psi^{\pm} &=-iM\bps^{\mp} e^{\vp^{\pm}},\qquad
\d\bps^{\pm}=iM\psi^{\mp}e^{\vp^{\pm}},\cr \d\bp\vp^{\pm}
&=iM\psi^{\pm}\bps^{\pm}e^{\vp\mp} +M^2e^{\vp^++\vp^-}.\ } }
The holomophic currents
\eqn\curr{ \eqalign{T &=-\d\vp^-\d\vp^+-{1\over
2}(\psi^-\d\psi^++\psi^+\d\psi^-)+{1\over 2}(\d^2\vp^++\d^2\vp^-),\cr
S^{\pm} &=-i\sqrt{2} (\psi^{\pm}\d\vp^{\pm}-\d\psi^{\pm}),\qquad
J=\d\vp^+-\d\vp^--\psi^-\psi^+,\ }} are conserved by $\bp
T=\bp S^{\pm}=\bp J=0$ on the equations of motion and similarly for
the antiholomorphic ones. One has to introduce also the generators
of $N=2$ supersymmetry $G^{\pm}$ and $\bar G^{\pm}$: \eqn\sus{
\eqalign{ G^{\pm}\vp^{\mp} &=i\sqrt{2}\psi^{\pm}, \hskip1cm
G^{\pm}\vp^{\pm}=0\cr \bar G^{\pm}\vp^{\mp}
&=i\sqrt{2}\bar\psi^{\pm}, \hskip1cm \bar G^{\pm}\vp^{\pm}=0\ }}
obeying the algebra: \eqn\alg{ \eqalign { \{G^+,G^-\} &=2\d,  \hskip
1cm \{G^{\pm},G^{\pm}\}=\{\bar G^{\pm},\bar G^{\pm}\}=0,\cr \{\bar
G^+,\bar G^-\} &=2\bp, \hskip1cm \{G,\bar G\}=0.\ }}
For the class
IIA fields only the chiral fields, $N_1^\o=e^{\o b\phi^+}$, 
has a classical limit. Their HEMs take the form:
$$
\bar G_{-\half}^+ G_{-\half}^+\phi^+N_1^\o =0,\qquad
\bar G_{-\half}^+ G_{-\half}^+\phi^-N_1^\o =B_1^\o N_1^{\o+1},
$$
where $B_1^{\o}=-2\pi i\mu b$ can be read from \bmo. In the
classical limit along with the analogous HEMs for class IIB
anti-chiral fields with $\o=0$, these become:
$$
\bar G^{\pm}G^{\pm}\vp^{\mp}=-2iMe^{\vp^{\pm}}.
$$
Together with \sus\ and the algebra \alg\ these relations encode the
equations of motion \em.

From the class I fields only the series $N_{1,n}^\o$ has a
classical limit, the simplest ``classical null-operators'' being:
$$
\eqalign{ D_{1,1}^{\o(cl)} &=\d-\half (\o+1)J+{1\ov {\o-1}}G^+G^-,
\cr D_{1,2}^{\o(cl)} &=\d^2-(\o+1)J\d-\half(\o +2)\d J+{1\ov
4}\o(\o+2)J^2+{2\ov {\o-2}}G^+G^-\d-{\o\ov {\o-2}}JG^+G^- - \cr
&-\half S^-G^++\half {{\o +2}\ov {\o-2}}S^+G^- . \ }
$$
It is easy to check, using the algebra \alg\ and the explicit form
of the currents \curr , that the classical expressions of the
corresponding null-vector conditions is:
$$
\eqalign{ D_{1,1}^{\o(cl)} e^{(\half (\o-1)\vp^+ -\half (\o
+1)\vp^-)} &=0,\cr D_{1,2}^{\o(cl)} e^{(\half (\o-2)\vp^+ -\half (\o
+2)\vp^-)} &=0 .\ }
$$
The same is of course  true also for $\bar D_{1,1}^{\o(cl)}$, $\bar
D_{1,2}^{\o(cl)}$. Then, with the help of \sus\ and the equations of
motion \em , we find that the classical HEMs then take the form:
$$
\eqalign{ \bar D_{1,1}^{\o(cl)} D_{1,1}^{\o(cl)} \vp^\pm e^{(\half
(\o-1)\vp^+-\half (\o+1)\vp^-)} &={{\o+1}\ov {\o-1}}M^2 e^{(\half
(\o+1)\vp^+ -\half (\o-1)\vp^-)},\cr \bar D_{1,2}^{\o(cl)}
D_{1,2}^{\o(cl)}\vp^\pm e^{(\half (\o-2)\vp^+-\half (\o+2)\vp^-)}
&=-2{{\o+2}\ov {\o-2}}M^4 e^{(\half (\o+2)\vp^+ -\half (\o-2)\vp^-)}.
\ }
$$
This is in a perfect agreement with \hemmn\ if we take into account
that the classical limit, $b\to 0$, of $B_{1,n}^\o=\bar B_{1,n}^\o$
from \bmno\ is:
$$
B_{1,n}^\o\to (-1)^{n+1}{{\o+n}\ov {\o-n}}n!(n-1)!\ b^{-1}(\pi\mu
b^2)^{2n}.
$$

To conclude, we presented relations among primary
fields, the higher equations of motion, in $N=2$ supersymmetric
Liouville field theory. We stress that, since in general the
null-vectors of this theory are unknown, our results \bmno\ and
\bmo\ should be understood as a proposal. Also, we were concerned in
this note with primary fields from the NS sector only. Since the
Ramond sector in $N=2$ SLFT is not very different, in particular the
degenerate fields fall into the same classes, we expect that very
similar HEMs hold for them too.

\vskip.5cm\no {\bf Acknowledgments}

We want to thank A. Belavin and V. Belavin for the interest in
this work. C.A. was supported in part by KRF-2007-313-C00150, WCU
Grant No.~R32-2008-000-101300. M. Stanshkov thanks IEU and CQUeST for the kind
hospitality. The work of M. Stanishkov is supported in part by NSFB
grant DO-02-257. M. Stoilov is supported by NSFB grant DO-02-288.

\listrefs

\bye